Propensity score matching in SPSS

Thoemmes, F.

University of Tübingen

Author Note

Felix Thoemmes, Center for Educational Science and Psychology, University of Tübingen, Europastr. 6, 72072 Tübingen, Germany, felix.thoemmes@gmail.com. Starting January 2012 Felix Thoemmes will be at the Department of Human Development, Cornell University, Ithaca, NY. The author would like to thank Philip Parker for valuable comments and Michael Becker for R computer code for dotplots.

Abstract

Propensity score matching is a tool for causal inference in non-randomized studies that allows for conditioning on large sets of covariates. The use of propensity scores in the social sciences is currently experiencing a tremendous increase; however it is far from a commonly used tool. One impediment towards a more wide-spread use of propensity score methods is the reliance on specialized software, because many social scientists still use SPSS as their main analysis tool. The current paper presents an implementation of various propensity score matching methods in SPSS. Specifically the presented SPSS custom dialog allows researchers to specify propensity score methods using the familiar point-and-click interface. The software allows estimation of the propensity score using logistic regression and specifying nearest-neighbor matching with many options, e.g., calipers, region of common support, matching with and without replacement, and matching one to many units. Detailed balance statistics and graphs are produced by the program.

## Propensity score matching in SPSS

Propensity score methods (Rosenbaum & Rubin, 1983) have seen a tremendous increase in use during the last couple of years (see Thoemmes & Kim, 2011), notably in the areas of education and evaluation research (e.g., Hong, & Raudenbush, 2005; Hughes, Chen, Thoemmes, & Kwok, 2010). Propensity score matching is a tool to adjust a treatment effect for measured confounders in non-randomized studies and is therefore an alternative to the commonly used regression adjustment (for an overview, see also Stuart, 2010). The logic behind propensity score methods is that balance on observed covariates is achieved through careful matching on a single score – the estimated propensity of selecting the treatment, or simply the propensity score. The propensity score is defined as the probability of receiving treatment based on measured covariates:

$$e(x) = P(Z=1 \mid \mathbf{X})$$

where *e(x)* is the abbreviation for propensity score, P a probability, Z=1 a treatment indicator with values 0 for control and 1 for treatment, the "|" symbol stands for conditional on, and **X** is a set of observed covariates. In other words, the propensity score expresses how likely a person is to select the treatment condition given observed covariates, e.g. person characteristics. This score is useful because it can be used to match participants from the treatment condition to participants from the control condition who have a very similar estimated propensity score. This matching process creates balance between treated and untreated participants on the propensity score and more importantly is also expected to create balance on the covariates that were used to estimate the propensity score. This balance property is a key aspect of propensity score methods because a balanced pre-test covariate cannot be a confounder anymore, i.e., cannot bias the treatment effect estimate. The balance that a randomized experiment is expected to create by design is here established through statistical matching.

Propensity score matching consists of several analytic steps:

1. Researchers select a set of pre-test covariates that are deemed important based on theoretical arguments. This step is critical as the credibility of the propensity score analysis hinges on the selection of proper covariates. Covariates of convenience (e.g. gender, age, income) are usually not sufficient (Shadish, Clark, & Steiner, 2008) instead researchers should strive to build a convincing case that no unobserved confounders are omitted. As an example Hong and Raudenbush (2005) present a study in which over 200 covariates were considered – one would be hard pressed to identify credible unobserved confounders in this case. Causal knowledge and assumptions about confounders can also be encoded in a directed acyclic graph (Pearl, 2000), which can then in turn be used by the researcher to select specific confounders for the estimation of the propensity score, however this approach is not yet widely adopted in the social sciences.

2. Based on this set of covariates the propensity score is estimated. This is often done using logistic regression in which the treatment assignment is used as the outcome variable, and the selected covariates as predictors. Covariates can be entered manually, based on theoretical and empirical considerations, or data mining approaches can be used (e.g. boosted regression trees, McCaffrey, Ridgeway, & Morral, 2004).

3. After estimation of the propensity score, the actual matching procedure commences. Matching can be performed in many different ways (see Stuart, 2010), but as Thoemmes and Kim (2011) note, most commonly straight-forward and simple techniques such as 1:1 nearest neighbor matching are used, meaning that a single treated participants is matched to a single untreated participant who has the most similar estimated propensity score. To ensure good matches, a caliper (maximum allowable difference between two participants) can be defined. If sample sizes of the treated and untreated participants vary greatly, one to many matching can be performed in which a single treated participant is matched to more than one untreated participant (see e.g., Ming, & Rosenbaum, 2000).

4. After matching is completed, a series of model adequacy checks should be performed. The main interest of the researcher is to check whether balance on the covariates has truly been achieved through the matching procedure. This can be done by comparing several statistics of the treatment and control group before and after matching, most often the standardized mean differences and the variance ratio. The standardized mean difference of covariates should be close to 0 after matching, and the variance ratio should be close to 1. In addition, bootstrapped Kolomgorov-Smirnov tests can be computed to examine equality of distribution of single covariates (Sekhon, 2011) and global imbalance measures (Iacus, King, & Porro, 2009) or multivariate significance tests (Hansen & Bowers, 2008) can be computed and assessed.

5. In a last step, the treatment effect is estimated in the matched subsample. Any statistical model (e.g., t-test, ANOVA, or more elaborate models, e.g. latent growth model [see Jackson, Thoemmes, Lüdtke, Jonkmann, & Trautwein (in press) for an example]) can be applied to the matched dataset. Recent research (Austin, 2011) argues that standard errors should be adjusted for the matched nature of the data, i.e. that one should compute a paired samples t-test to examine mean differences between two groups, however this is an issue that is still debated (see e.g. Stuart, 2010).

Currently, several programs exist that perform these steps in a propensity score analysis, however they are primarily written in R (Ho, Imai, King, & Stuart, 2007) or consist of special macros in Stata or SAS. A thorough implementation in SPSS is lacking and the current paper attempts to fill this gap. Given that many applied researchers in psychology, education, and other social sciences still primarily use SPSS as their main data analysis tool and are often not intimately familiar with R, this seems to be a useful addition. The program "psmatching" is written as a so-called custom dialog in SPSS and works with versions 18, 19 and 20 of SPSS. The custom dialog provides the user with the familiar point-and-click interface and generates SPSS syntax that can be pasted and modified if needed. The "psmatching" program performs all analyses in R through the SPSS R-Plugin and makes use of newly written R code by the author of this manuscript and some R packages written by other researchers. In particular the following packages are invoked: "MatchIt" by Ho, Imai, King, & Stuart (2007), "RItools" by Bowers, Fredrickson, & Hansen (2010), and "cem" by Iacus, King, & Porro (2009). All of those packages are explicitly acknowledged in the program, whenever they are used. The "MatchIt" package is used in a central way, and the author explicitly acknowledges that analytic choices are guided by the work of the authors of the "MatchIt" package (Ho, Imai, King, & Stuart, 2007).

### SPSS custom dialog

Appendix A explains where to obtain and how to install the custom dialog to perform propensity score matching. Once installed, the SPSS custom dialog becomes part of the Analyze Menu in SPSS and allows researchers to estimate propensity scores for a binary treatment variable from a set of specified covariates and subsequently perform matching. Figure 1 shows the interface and available analysis options.

Figure 1. Screenshot of the SPSS PS Matching Custom Dialog.

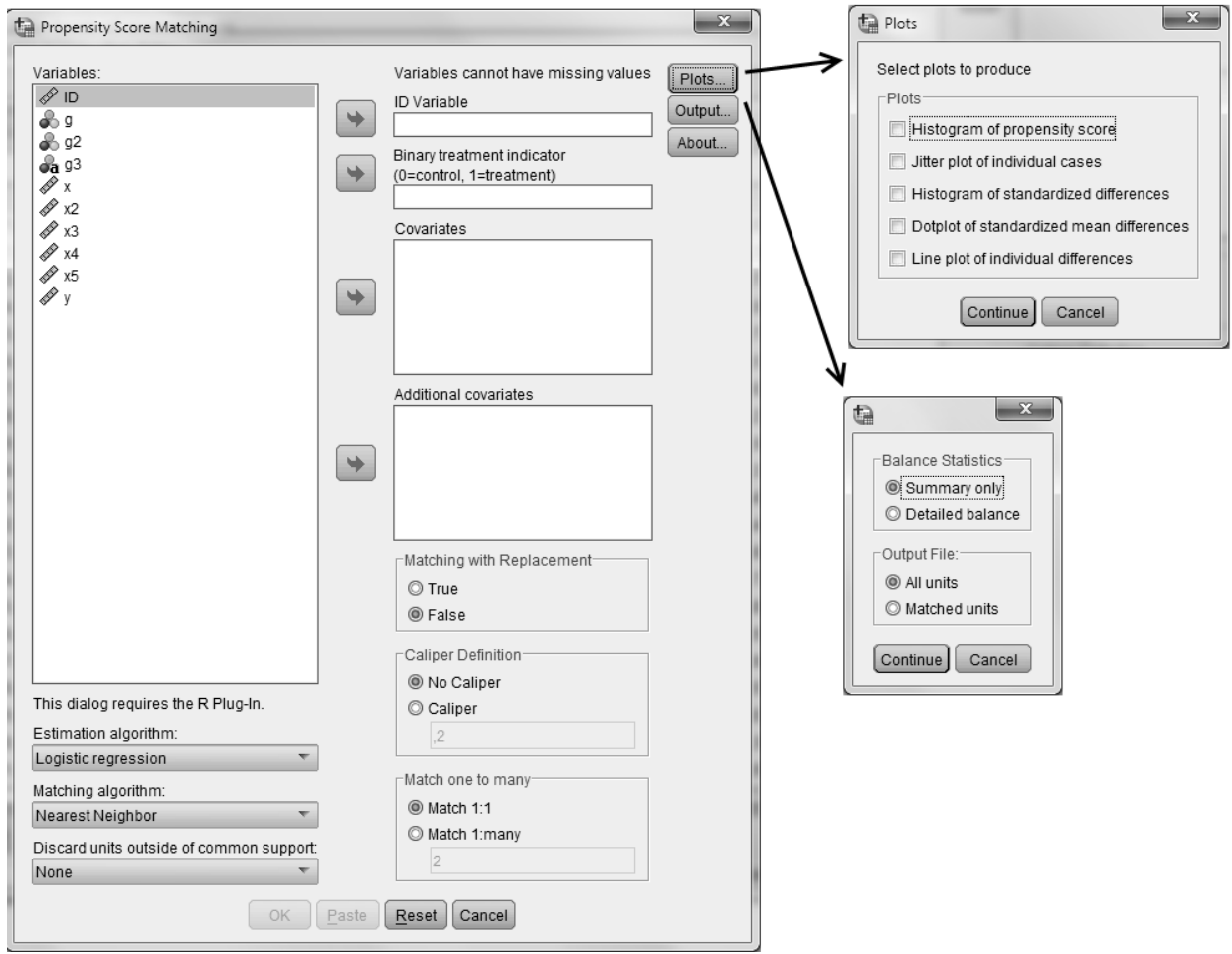

Outlined below are the analytic choices that can be made by the researchers in the program and the underlying statistical models that are used.

**Selection of treatment variable and covariates**

The SPSS custom dialog accepts a single treatment variable and a theoretically unlimited number of covariates as input. The treatment variable has to be binary with the control condition coded 0 and the treatment condition coded 1. The covariates are used to predict treatment assignment using logistic regression, specified as:

$$Ln\left[\frac{P(Z=1|X_1,...,X_j)}{1-P(Z=1|X_1,...,X_j)}\right] = \beta_0 + \sum_{j=1}^{p} \beta_j X_j \qquad (1)$$

where Z is the binary variable indicating treatment or control condition, and $X_1$ to $X_j$ are all covariates that are being used to predict group membership of the binary treatment variable. Note that interactions and quadratic terms can be entered manually after the user created them in the SPSS dataset. The predicted values of this equation are the estimated propensity scores. Note that other statistical models could be used to estimate propensity scores, e.g. discriminant analysis, regression trees, or free-from regressions, e.g. lowess smoothers, but currently the software only implements the logistic regression model.

The dialog also allows for additional covariates to be entered that are not used in the estimation of the propensity score but are evaluated on their balance nonetheless. This is especially helpful when a covariate is assumed to be not related to treatment selection and is excluded from the estimation model. As a general rule, balance should be examined on all pre-test covariates, regardless of whether they are entered in the estimation of the propensity score or not.

**Matching algorithm**

Currently, the SPSS custom dialog implements nearest neighbor matching, a simple routine to find matches in two groups that is based on a greedy matching algorithm that sorts the observations in the treatment group by their estimated propensity score and matches each unit sequentially to a unit in the control group that has the closest propensity score, i.e. is the nearest neighbor of this unit. Several options are available to improve or fine-tune this matching: the custom dialog allows the user to select whether units outside the area of common support (defined as the region of the distributions of estimated propensity scores in the treatment and control group for which units in both groups are observed) should be discarded or included in the matching algorithm. Specifically, the option allows users to discard no units whatsoever, discard only units from the treatment group that fall outside the region of common support, discard only units from the control group that fall outside the region of common support, or all units that fall outside the region of common support. Excluding units outside the area of common support can improve balance on covariates, and can avoid extrapolation to units in one group that were so dissimilar on their covariates that no comparable units in the other group were found. Note that discarding units changes the estimate of interest and the average causal effect is no longer identified, but rather a local causal effect for units that were comparable enough to warrant effect estimation without extrapolation.

The user can also specify matching with replacement, in which a single unit in the control group can be reused to be matched to more than one unit in the control group. Usually, this reduces the overall imbalance between the two groups, because the closest possible unit in the control group can be used for matching, even if this unit also has been used for a different match. However, under some circumstances very few units can be repeatedly matched to units in the treatment group, making the estimate of the treatment effect somewhat dependent on these highly reused units.

Another option to fine-tune the matching is to request ratio matching in which a specified number of control units can be matched to a single treatment unit, e.g. a 2:1 ratio matching means that up to 2 control units (with similar estimated propensity score) can be matched to a single unit in the treatment group. This is usually helpful when the sample sizes of the two groups differ substantially, and 1:1 matching would discard many units that could be potentially matched. Ming and Rosenbaum (2000) showed that ratio matching can be beneficial, but that the advantage in terms of balance and precision are usually reached with 5 matches to a single unit. Both under ratio matching and matching with replacement, weights are generated to account for the fact that certain units are matched with more than one unit. The exact generation of the weights is described in more detail by Ho, Imai, King, and Stuart (2007), but briefly described, all treated matched units receive a weight of 1 and all unmatched units receive a weight of 0. Control units under regular 1:1 nearest neighbor matching also receive a weight of 1. When one to many matching is employed, control units are given an initial weight proportionate to the number of treated and control units, which is then rescaled so that the sum of weights equals the number of matched treated units. In the case of matching with replacement, weights of control units that were reused are summed across all matches in which the control unit was used.

      Finally, the user can specify a so-called caliper that prevents "bad" matches, i.e., matches of units that have very dissimilar estimated propensity scores and hence are most likely imbalanced on their covariates as well. A caliper is a maximum distance that two units can be apart from each other (on their estimated propensity scores) and is defined in units of standard deviations of the logit of the estimated propensity score. Defining a small caliper will usually result in better balance at the expense of finding fewer units that can be successfully matched. Conversely, a large caliper will retain more matches, but some of them will be slightly imbalanced, and might yield a larger bias in the estimation of the treatment effect. Whenever a

caliper is defined each treated unit will be matched to one (or more, depending on the options chosen by the user) control unit that is randomly drawn out of all control units that fall within the caliper.

**Output options**

The SPSS custom dialog provides the user with detailed output including different balance statistics and graphics to assess achieved balance. First, sample sizes for both groups are reported for pre- and post-matching samples. In addition, units that were not matched are further divided into units that were discarded due to falling outside the region of common support or discarded because no proper match could be found. For unweighted data, an overall imbalance $\chi^2$ test is provided. The test was developed by Hansen and Bowers (2008) and is defined as:

$$d^2(\mathbf{z}; x_1, \ldots, x_j) := [d(\mathbf{z}, x_1), \ldots, d(\mathbf{z}, x_j)] \times \left( Cov \begin{bmatrix} d(\mathbf{Z}, x_1) \\ \vdots \\ d(\mathbf{Z}, x_j) \end{bmatrix} \right)^{-} \times \begin{bmatrix} d(\mathbf{z}, x_1) \\ \vdots \\ d(\mathbf{z}, x_j) \end{bmatrix} \qquad (2)$$

where $d$ is a group difference (mean difference for continuous variables) on variables denoted $x$, based on groups denoted by $z$. The middle expression on the right hand side is the inverse of the covariance matrix of group differences. This test statistic, which is related to the well-known Hotelling's $T^2$ statistic, assesses simultaneously whether any variable or any linear combination of variables was significantly unbalanced after matching. The test examines all covariates that were used to estimate the propensity score and all variables that were defined as additional covariates by the user.

A second multivariate overall imbalance measure is provided in the form of $\mathcal{L}_1$ developed by Iacus, King, and Porro (2009). The $\mathcal{L}_1$ measure is based on an automatic coarsening of all variables into bins and then comparing differences in frequencies of all cells of a multivariate contingency table of the control and treatment group on all binned (discretized) variables. The $\mathcal{L}_1$ measure is defined as

$$\mathcal{L}_1 = \frac{1}{2}\sum_{\ell_1\ldots\ell_j} |t_{\ell_1\ldots\ell_k} - c_{\ell_1\ldots\ell_k}| \tag{3}$$

where $\ell$ is the frequency of a given cell, indexed by 1 to k, in the multivariate contingency table, for either the treatment group $t$ or control group $c$. The test simultaneously assesses the full joint distribution (based on the discretization) of all covariates, meaning that higher order polynomials and interactions, are tested as well. The $\mathcal{L}_1$ measure is bounded by 0 (perfect balance) and 1 (complete separation in the cross-tabulation). There is no direct cutoff value that indicates good or bad balance, but rather the $\mathcal{L}_1$ measure must be compared to the unmatched solution or other alternative matching solutions. A desirable situation is that the $\mathcal{L}_1$ measure is smaller in the matched than in the unmatched sample and that the chosen matching solution has a low $\mathcal{L}_1$ measure compared to other solutions that may have been tried by the researcher. More details on the use and interpretation of the $\mathcal{L}_1$ measure is given by Iacus, King, and Porro (2011). The SPSS dialog reports the measure for both before and after matching and allows for a convenient comparison of this total relative imbalance measure. The custom dialog ensures that the discretization of the matched and unmatched groups is the same, therefore allowing a meaningful comparison of the two $\mathcal{L}_1$ measures. The actual binning of variables is based on an automatic algorithm implemented in the cem package (Iacus, King, & Porro, 2009). Note that the algorithm tends to have relatively fine binning and that therefore $\mathcal{L}_1$ values close to 1.0 are relatively common when many covariates are considered. The test is again based on all covariates that were specified by the user.

Besides the two multivariate tests, univariate tests are provided for each covariate, each quadratic term of the covariates, and every possible interaction. For each of these terms, the group means of treated and untreated participants are reported, alongside with the standard deviation of the control group, and a standardized mean difference, defined as the mean difference between the groups divided by the standard deviation of the control group. These statistics are reported for the pre- and post-matching samples. These lists can get rather long, which makes it very difficult to spot large differences. The dialog therefore allows suppressing these tables and instead presents only a condensed table of large imbalances, defined as all terms that have a standardized mean difference larger than .25. This condensed table is also sorted by the magnitude of the imbalance, making it easy for users to spot covariates or quadratic and interaction terms that need additional balancing, e.g. through re-specification of the propensity score.

In addition to these numerical balance measures, a total of five different diagnostic plots can be requested. Three of these plots are automatically generated by the MatchIt package (Ho, Imai, King, & Stuart, 2007) and the remaining two plots are a novel addition of the SPSS custom dialog. First, a histogram of the propensity scores in both groups before and after matching can be requested. The histograms are by default always overlaid with a kernel density estimate. This graph allows for a visual inspection of the similarity of the propensity score distributions after matching and also an assessment of the area of common support. Tail regions of histograms or kernel density estimates that are non-overlapping in the distributions of the propensity scores in treatment and control group are an indication of insufficient overlap. The second graph that can be requested is a dotplot of individual propensity scores of units in the control and treatment group, either matched or unmatched. This graph provides similar information as the histograms with overlaid kernel density estimates but plots individual units, which makes it easier to visualize individual matches. In the case of one to many matching, weights are represented as dots of differing sizes. The third plot that can be requested is a pair of histograms (with overlaid kernel density estimates) that show the standardized differences of all terms (covariates, quadratic term, interactions) before and after matching. The two histograms are forced to be on the same scales to afford an easy comparative view of the magnitude of differences before and after matching. Also, the histograms make it easy to discern whether standardized differences after matching are centered on zero and that no systematic differences still exist after matching. The fourth plot is another dotplot that displays the magnitude of the standardized differences before and after matching for each covariate. This plot mainly serves as an illustration of descriptive information. Covariates that appear in the dotplot are in the same order as they are encountered in the dataset. Finally, a fifth plot can be requested that shows standardized mean differences before and after matching in the form of a parallel line plot. Standardized differences that increase after matching are bolded. Examples of all graphs are given in the applied example below.

      As a last feature of the SPSS custom dialog, the user can request to append all estimated propensity scores and weights to a new copied dataset, or create a new dataset that contains only units that were successfully matched and appends propensity scores and weights for these units only. Both datasets are created automatically in SPSS and after turning on weights in SPSS, the user can run all desired analyses on the matched sample.

## Illustrative Example

We illustrate the use of the propensity score matching SPSS custom dialog on an applied example that is conducted using simulated data. The simulated data are based on the TOSCA study (Köller, Watermann, Trautwein, & Lüdtke, 2004) – a large longitudinal study in Germany assessing various psychosocial and academic outcomes of young adults. We identified a number of key variables in the TOSCA dataset, and using the covariance matrix and means of these variables, we simulated 4148 cases (this number was identical to the number of cases in the actual TOSCA dataset). By using a simulated dataset (as opposed to the actual data) we avoid issues of missing data and data confidentiality. The question of interest was whether the decision of young adults to continue living with their parents after high-school graduation compared to moving out of the household has any impact on certain personality characteristics assessed one year later after this decision. In this example, only the impact of moving out from the parents' household on neuroticism is assessed. Clearly, observed differences in neuroticism one year after moving out may be due to other pre-existing differences, such as initial neuroticism levels, or other personality differences that existed before the decision is made to move out. If pre-existing differences are also related to the outcome variable (neuroticism measured one year later), then the estimation of the true causal effect of moving out will be biased due to confounding. Our example serves mainly illustrative purposes and we focus on the presentation of the method, rather than making any substantive claims about the plausibility of observed effects.

The simulated data consisted of a total of 4148 young adults measured at baseline. Observed covariates at baseline included current neuroticism level, gender, type of high school (vocational track, academic track), age, IQ, socio-economic status, intentions to go to a University, and problems with parents. One year later, students were again measured on all variables. Of the initial 4148 participants, 2636 stayed with their parents, and 1512 moved out of their parents' household.

The unadjusted estimate of the effect of moving out on neuroticism measured one year later was statistically significant ($t(4146) = -12.39, p < .001$). Young adults who decided to stay with their parents exhibited lower mean neuroticism scores ($M = 2.96$) than young adults who chose to leave the parents' household ($M = 3.16$) and the standardized mean difference was $d = -.41$. This effect estimate does not necessarily represent the true causal effect of moving out, because as mentioned earlier, many covariates could be potential confounders. As an example, the amount of problems with parents for young adults who stayed at home ($M = 3.37$) was significantly lower ($t(4146) = -6.19, p < .001$) than those of young adults that moved away from home ($M = 3.67$). Simultaneously, amount of problems with parents was also slightly, but significantly related to the outcome variable neuroticism ($r = .05, p = .004$), making it a variable that can bias the estimated effect of moving out.

To control for these confounding influences, we conducted a propensity score analysis using the SPSS custom dialog. In a first step the propensity score, i.e. the probability of moving out of the parents' household was estimated using logistic regression. For this particular example we used all covariates mentioned previously. After estimation of the propensity score, we matched participants using a simple 1:1 nearest neighbor matching. In order to exclude bad matches (in a sense that the estimated propensity score from two matched units are very different from each other), we imposed a caliper of .15 of the standard deviation of the logit of the propensity score. After matching we examined the balance of all observed covariates, interactions among all covariates, and quadratic terms of all covariates. Nearly no imbalances remained as assessed through univariate and multivariate tests. The largest remaining standardized difference after matching was found at the interaction between socio-economic status and problems with parents with a value of d = .09, a rather small standardized difference.

Figure 2. Dotplot of standardized mean differences (Cohen's *d*) for all covariates before and after matching.

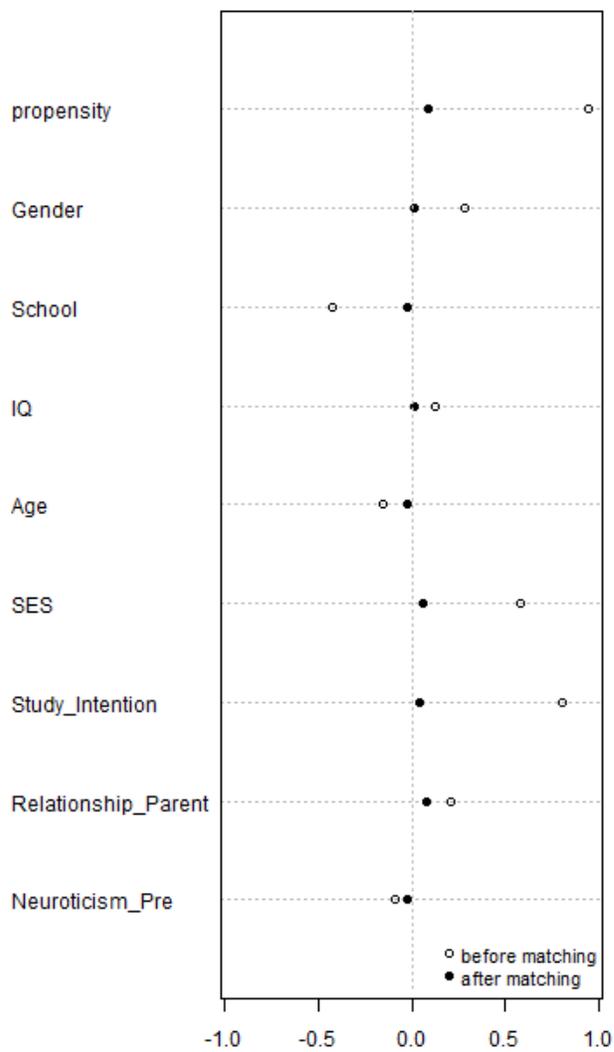

The overall $\chi^2$ balance test was also not significant, $\chi^2(8) = 13.89$, p = .09. The $\mathcal{L}_1$ measure was larger in the unmatched sample (.938) than in the matched sample (.919), also indicating that matching improved overall balance.

All five diagnostic plots were produced and are displayed in Figures 2 to 6. Figure 2 shows a dotplot of covariate balance in terms of standardized mean differences for all individual covariates, Figure 3 shows the actual propensity score distributions of both groups before and after matching overlaid with a kernel density estimate, Figure 4 shows a lineplot of standardized differences before and after matching, Figure 5 shows histograms with overlaid kernel density estimates of standardized differences before and after matching, and finally Figure 6 displays a dotplot of individual units in the dataset and whether they were matched or discarded.

Figure 3. Distribution of propensity scores of young adults staying with their parents ("treated") and young adults moving out ("control") before and after matching with overlaid kernel density estimate. Graph was produced using modified routines of the MatchIt package.

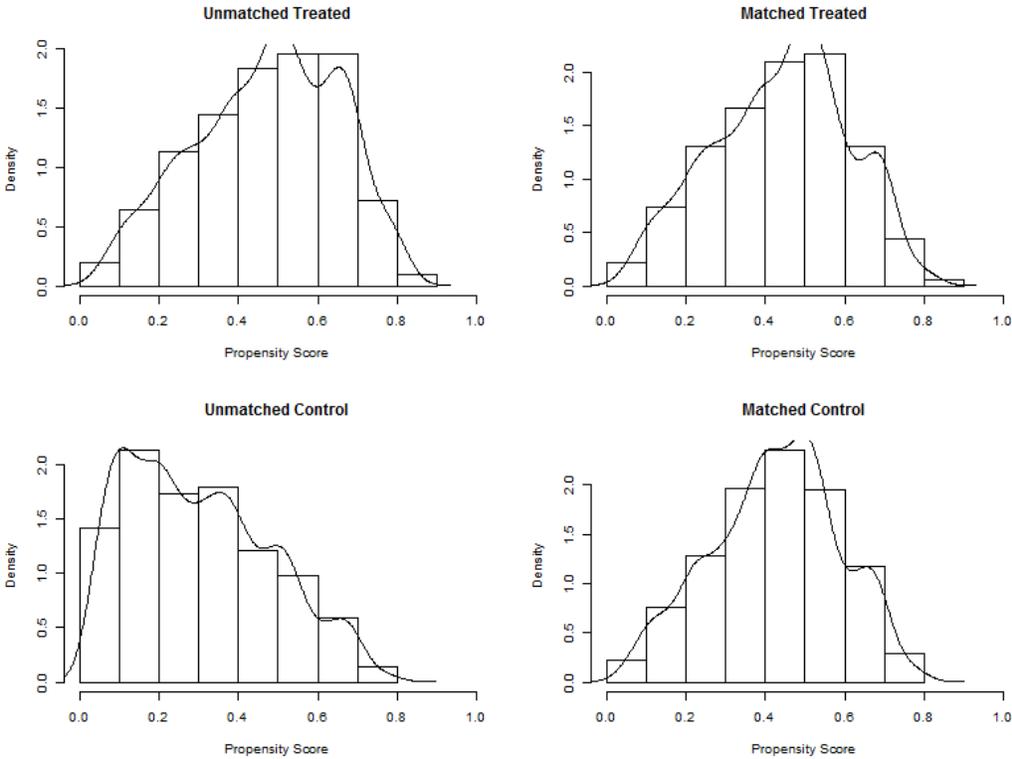

Figure 4. Lineplot of standardized differences before and after matching. Graph was produced using routines of the MatchIt package.

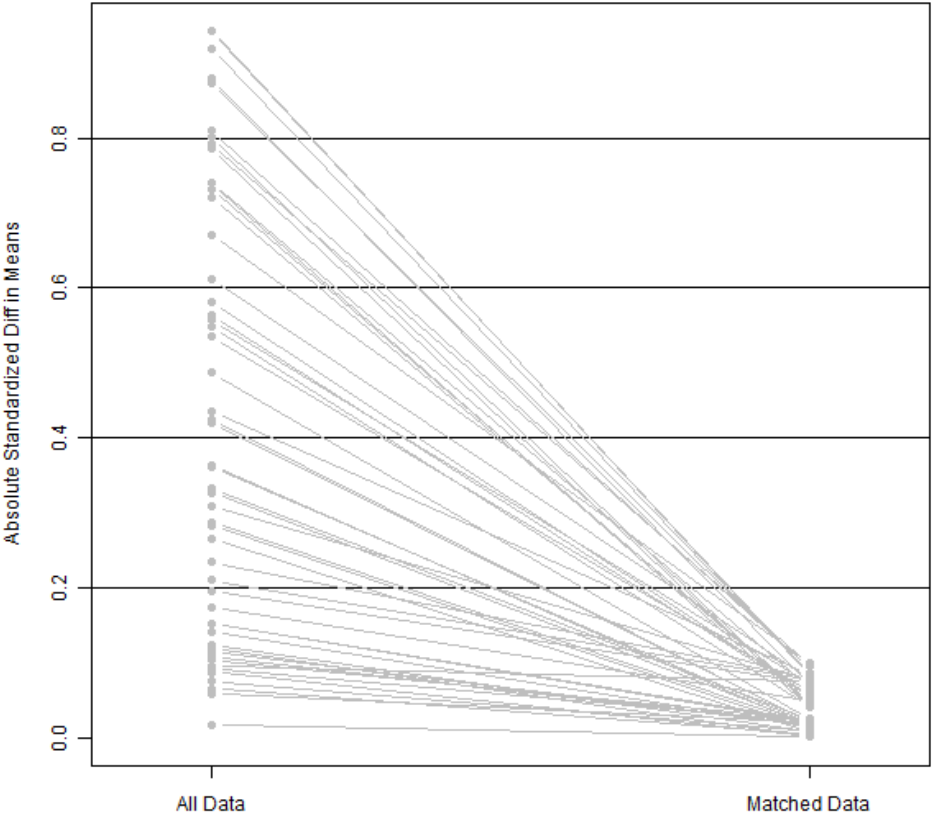

Figure 5. Histograms with overlaid kernel density estimates of standardized differences before and after matching.

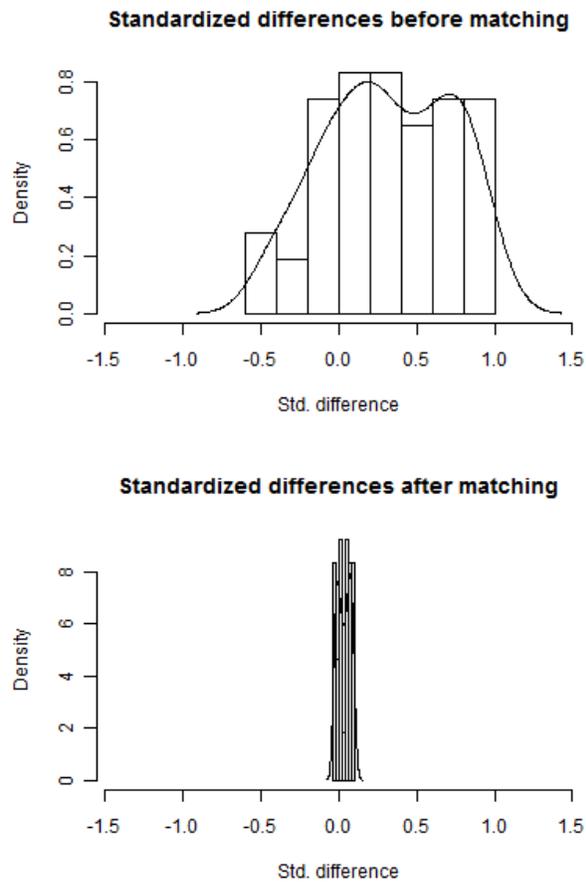

Figures 2-5 show that covariate balance was massively improved in the matched sample, while Figures 3 and 6 show that the region of common support spanned almost the entire distribution of the propensity score and that only in extreme tail regions no appropriate matches could be found.

Figure 6. Dotplot of individual young adults in either matched or unmatched groups. Graph was produced using routines from the MatchIt package.

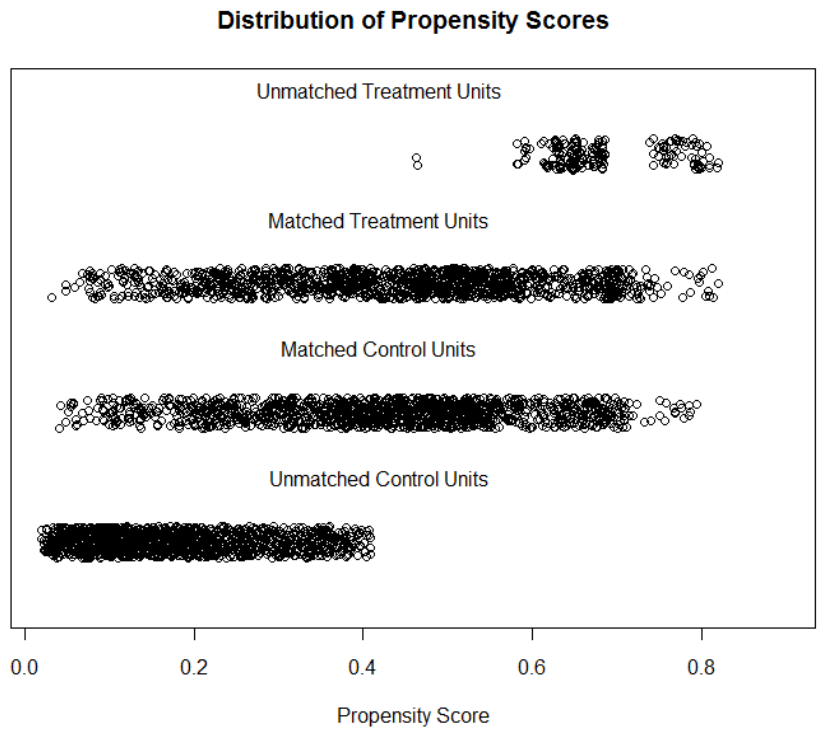

The matched sample now included a total of 2676 young adults, evenly distributed in the two groups ("moving out" versus "staying at home"). The adjusted estimate of the effect of moving out on neuroticism in the matched sample was (very slightly) not significant anymore ($t(2674) = -1.93$, $p = .053$). This drop in significance was also not purely a function of the smaller sample size but also the overall magnitude of the effect diminished to a decreased standardized mean difference of $d = -.11$, from previously $d = -.41$, in the unmatched sample. Young adults who stayed at home had only slightly lower neuroticism scores ($M = 3.09$) than young adults who decided to move out of their parents' home ($M = 3.13$) in the matched sample. The unadjusted effect that was highly significant before propensity score matching therefore shrunk in absolute magnitude and also fell slightly short of the conventional significance threshold of .05, indicating that some of the effect that was originally observed was simply due to confounding influences of covariates and not due to causal effects of moving out.

### Limitations and future developments

The current implementation of propensity score matching in SPSS is limited in regards to several aspects. First, some matching algorithms are not yet implemented. Examples are optimal matching (Hansen & Klopfer, 2006), full matching (Rosenbaum, 1991), coarsened exact matching (Iacus, King, & Porro, 2011), genetic matching (Sekhon, 2011), interval matching (Pan, & Bai, 2011), or entropy matching (Hainmueller, 2011). In addition, some more advanced estimation algorithms, e.g. boosted regression trees (McCaffrey, Ridgeway, & Moral, 2004), or generalized additive functions are not yet implemented. Work is being conducted to address these limitations and include other matching and estimation techniques. Further, as indicated in the software itself, we have not yet implemented the use of the Hansen & Bowers (2008) global imbalance test when data are weighted, e.g. every time when one to many matching schemes are used. In addition, there are several other balance statistics that could potentially be reported, e.g. variance ratios of variables in the matched sample of treated and untreated units, or Kolmogorov-Smirnov tests that examine the equality of the full distributions of variables in the groups of treated and untreated subjects. The $\mathcal{L}_1$ measure is currently only computed for one particular binning of the variables, which occasionally results in $\mathcal{L}_1$ values that are very close to 1. Allowing adjustment of the binning or examining plots over a wide range of different binnings (as already implemented in the cem package (Iacus, King, & Porro, 2009)) would be desirable.

The propensity score matching dialog currently does not handle missing data and will report, upon encountering missingness, that missing values on covariates or the treatment variable is not admissible. One possible solution is to generate multiple imputations and repeat the analysis on each of the imputed datasets (see, e.g. Hill, 2004). Also, currently only binary treatment variables are supported and multi-valued categorical, ordinal, or continuous treatment variables cannot be analyzed. A solution for examining effects of multi-valued treatments suggested by Rubin (1997) is to compare two treatment conditions at a time and examine all interesting pairwise comparisons that result from splitting up a multi-valued categorical treatment.

## Discussion

The adjustment of observed effects in non-randomized studies is a critical part of data analysis, because confounding influences of covariates can bias effect estimates. Propensity score methods offer a principled approach to deal with this type of confounding bias. Through efficient matching, balance is created on the covariates and their confounding effect can be minimized or entirely removed. Given a theoretically sound set of potential confounders, researchers are able to reduce bias substantially and, if they are able to convincingly demonstrate the absence of unobserved confounders, can theoretically estimate an unbiased causal effect of a treatment – usually a very desirable quantity for applied researchers.

Currently propensity score methods are increasingly used but they do not seem to enjoy widespread use in psychological and educational sciences. We believe one impediment was the lack of software options in statistical programs that most social scientists are accustomed to. Our goal was to supply such a tool and we would hope to see it put to good use!

PROPENSITY SCORE MATCHING IN SPSSHughes, J. N., Chen, Q., Thoemmes, F., & Kwok, O. (2010). An Investigation of the Relationship Between Retention in First Grade and Performance on High Stakes Tests in Third Grade. *Educational Evaluation and Policy Analysis, 32*, 166-182.

Iacus, S. M., King, G., & Porro, G. (2009). CEM: Coarsened exact matching software. *Journal of Statistical Software, 30*, 1-27.

Iacus, S. M., King, G., & Porro, G. (2011). Causal Inference without Balance Checking: Coarsened Exact Matching. *Political Analysis*, *20,* 1-24.

Jackson, J., Thoemmes, F., Lüdtke, O., Jonkmann, K., & Trautwein, U. (in press). Military training and personality trait development: Does the military make the man or does the man make the military? *Psychological Science.*

Köller, O., Watermann, R., Trautwein, U., & Lüdtke, O. (2004). *Wege zur Hochschulreife in Baden-Württemberg: TOSCA – Eine Untersuchungan allgemein bildenden und beruflichen Gymnasien.* VS Verlag für Sozialwissenschaften. Wiesbaden, Germany.

McCaffrey, D. F., Ridgeway, G., & Morral, A. R. (2004). Propensity Score Estimation With Boosted Regression for Evaluating Causal Effects in Observational Studies. *Psychological Methods, 9,* 403-425.

Ming, K. & Rosenbaum, P. (2000). Substantial gains in bias reduction from matching with a variable number of controls. *Biometrics, 56*, 118-124.

Pan, W., & Bai, H. (2011, August). *Interval Matching: Propensity Score Matching Using Case-Specific Bootstrap Confidence Intervals.* Paper presented at the annual Joint Statistical Meeting, Miami Beach, Florida.

Pearl, J. (2000). *Causality: models, reasoning, and inference.* Cambridge University Press: New York.

PROPENSITY SCORE MATCHING IN SPSS

Appendix A.

Installation instructions for SPSS R plug-in (SPSS R Essentials) and Custom Dialog "PS Matching"

1.) Determine which version of SPSS you are running and install the correct version of R and the correct version of the SPSS R plug-in (SPSS R Essentials). You can find out which version of SPSS you have installed by clicking "Help → About". Currently, SPSS provides the R Essential tool for SPSS 18, 19 and 20. If you have SPSS 19 installed, you will need to install R 2.8.1, and if you have SPSS 20 installed, you will need to install R 2.12.0. Other versions of R will not work, even if they are newer (e.g., R 2.13). You can find older releases of R on the website http://cran.r-project.org/ and then clicking on "Download R for Windows" (or your alternative operating system), followed by a click on "base", "previous releases", and finally the specific R version that you would like to download. After installation of R, obtain the SPSS R plug-in (SPSS R Essentials) The R plug-in can currently be downloaded from http://www.ibm.com/developerworks/spssdevcentral or directly from IBM at https://www14.software.ibm.com/webapp/iwm/web/preLogin.do?source=swg-tspssp. You may have to register for a free account on IBM.com to download the R plug-in. Each version of SPSS and each operating system (Windows 32bit, 64bit, etc.) has its own version of the plug-in. The custom dialog file (psmatching.spd) that contains the actual code to perform propensity score matching in SPSS can be downloaded from

http://sourceforge.net/projects/psmspss/files/.

2.) After successfully installing the correct versions of R and the SPSS R-plug-in it is recommended to test whether both components are working properly. Open SPSS and write in a syntax box the following commands:

```
BEGIN PROGRAM R.
x <- "R plug-in is working properly"
x
END PROGRAM.
```

If your output takes on the following form, all components are working correctly.

```
BEGIN PROGRAM R.
x <- "R plug-in is working properly"
x
END PROGRAM.
[1] " R plug-in is working properly"
```

3.) Open SPSS in administrator model (this is especially important for Windows 7 users). Do so, by right-clicking the SPSS Icon and choose "Run as administrator". In SPSS navigate to "Utilities → Custom Dialogs → Install Custom Dialog" and choose the .spd file that contains the "SPSS Propensity Score Matching" Custom Dialog. SPSS should now display an additional icon to conduct propensity score matching.